# Citation Statistics
## A Report from the International Mathematical Union (IMU) in Cooperation with the International Council of Industrial and Applied Mathematics (ICIAM) and the Institute of Mathematical Statistics (IMS)

Robert Adler, John Ewing and Peter Taylor

*Robert Adler, Faculty of Electrical Engineering, Faculty of Industrial Engineering and Management, Technion–Israel Institute of Technology. John Ewing is President, American Mathematical Society. Peter Taylor, Department of Mathematics and Statistics, University of Melbourne. The authors comprised the Joint IMU/ICIAM/IMS-Committee on Quantitative Assessment of Research with John Ewing acting as chair.*

[1]Discussed in 10.1214/09-STS285A, 10.1214/09-STS285B, 10.1214/09-STS285C and 10.1214/09-STS285D; rejoinder at 10.1214/09-STS285REJ.

http://www.mathunion.org/publications/report/citationstatistics/

## EXECUTIVE SUMMARY

This is a report about the use and misuse of citation data in the assessment of scientific research. The idea that research assessment must be done using "simple and objective" methods is increasingly prevalent today. The "simple and objective" methods are broadly interpreted as *bibliometrics*, that is, citation data and the statistics derived from them. There is a belief that citation statistics are inherently more accurate because they substitute simple numbers for complex judgments, and hence overcome the possible subjectivity of peer review. But this belief is unfounded.

- Relying on statistics is not more accurate when the statistics are improperly used. Indeed, statistics can mislead when they are misapplied or misunderstood. Much of modern bibliometrics seems to rely on experience and intuition about the interpretation and validity of citation statistics.
- While numbers appear to be "objective," their objectivity can be illusory. The meaning of a citation can be even more subjective than peer review. Because this subjectivity is less obvious for citations, those who use citation data are less likely to understand their limitations.
- The sole reliance on citation data provides at best an incomplete and often shallow understanding of research—an understanding that is valid only when reinforced by other judgments. *Numbers are not inherently superior to sound judgments.*

Using citation data to assess research ultimately means using citation-based statistics to rank things—journals, papers, people, programs, and disciplines. The statistical tools used to rank these things are often misunderstood and misused.

- For journals, the impact factor is most often used for ranking. This is a simple average derived from the distribution of citations for a collection of articles in the journal. The average captures only a small amount of information about that distribution, and it is a rather crude statistic. In addition, there are many confounding factors when judging journals by citations, and any comparison of journals requires caution when using impact factors. Using the impact factor alone to judge a journal is like using weight alone to judge a person's health.
- For papers, instead of relying on the actual count of citations to compare individual papers, people frequently substitute the impact factor of the journals in which the papers appear. They believe that higher impact factors must mean higher citation counts. But this is often *not* the case! This is a pervasive misuse of statistics that needs to be challenged whenever and wherever it occurs.

- For individual scientists, complete citation records can be difficult to compare. As a consequence, there have been attempts to find simple statistics that capture the full complexity of a scientist's citation record with a single number. The most notable of these is the h-index, which seems to be gaining in popularity. But even a casual inspection of the h-index and its variants shows that these are naïve attempts to understand complicated citation records. While they capture a small amount of information about the distribution of a scientist's citations, they lose crucial information that is essential for the assessment of research.

The validity of statistics such as the impact factor and h-index is neither well understood nor well studied. The connection of these statistics with research quality is sometimes established on the basis of "experience." The justification for relying on them is that they are "readily available." The few studies of these statistics that were done focused narrowly on showing a correlation with some other measure of quality rather than on determining how one can best derive useful information from citation data.

We do not dismiss citation statistics as a tool for assessing the quality of research—citation data and statistics can provide some valuable information. We recognize that assessment must be practical, and for this reason easily derived citation statistics almost surely will be part of the process. But citation data provide only a limited and incomplete view of research quality, and the statistics derived from citation data are sometimes poorly understood and misused. Research is too important to measure its value with only a single coarse tool.

We hope those involved in assessment will read both the commentary and the details of this report in order to understand not only the limitations of citation statistics but also how better to use them. If we set high standards for the conduct of science, surely we should set equally high standards for assessing its quality.

> **From the committee charge**
> The drive towards more transparency and accountability in the academic world has created a "culture of numbers" in which institutions and individuals believe that fair decisions can be reached by algorithmic evaluation of some statistical data; unable to measure quality (the ultimate goal), decision-makers replace quality by numbers that they can measure. This trend calls for comment from those who professionally "deal with numbers"—mathematicians and statisticians.

## INTRODUCTION

Scientific research is important. Research underlies much progress in our modern world and provides hope that we can solve some of the seemingly intractable problems facing humankind, from the environment to our expanding population. Because of this, governments and institutions around the world provide considerable financial support for scientific research. Naturally, they want to know their money is being invested wisely; they want to assess the quality of the research for which they pay in order to make informed decisions about future investments.

This much isn't new: People have been assessing research for many years. What *is* new, however, is the notion that good assessment must be "simple and objective," and that this can be achieved by relying primarily on metrics (statistics) derived from citation data rather than a variety of methods, including judgments by scientists themselves. The opening paragraph from a recent report states this view starkly:

> It is the Government's intention that the current method for determining the quality of university research—the UK Research Assessment Exercise (RAE)—should be replaced after the next cycle is completed in 2008. Metrics, rather than peer-review, will be the focus of the new system and it is expected that bibliometrics (using counts of journal articles and their citations) will be a central quality index in this system (Evidence Report, 2007, page 3).

Those who argue for this simple objectivity believe that research is too important to rely on subjective judgments. They believe citation-based metrics bring clarity to the ranking process and elimi-nate ambiguities inherent in other forms of assessment. They believe that carefully chosen metrics are independent and free of bias. Most of all, they believe such metrics allow us to compare all parts of the research enterprise—journals, papers, people, programs and even entire disciplines—simply and effectively, without the use of subjective peer review.

But this faith in the accuracy, independence and efficacy of metrics is misplaced.

- First, the accuracy of these metrics is illusory. It is a common maxim that statistics can lie when they are improperly used. The misuse of citation statistics is widespread and egregious. In spite of repeated attempts to warn against such misuse (e.g., the misuse of the impact factor), governments, institutions, and even scientists themselves continue to draw unwarranted or even false conclusions from the misapplication of citation statistics.
- Second, sole reliance on citation-based metrics replaces one kind of judgment with another. Instead of subjective peer review one has the subjective interpretation of a citation's meaning. Those who promote exclusive reliance on citation-based metrics implicitly assume that each citation means the same thing about the cited research—its "impact." This is an assumption that is unproven and quite likely incorrect.
- Third, while statistics are valuable for understanding the world in which we live, they provide only a partial understanding. In our modern world, it is sometimes fashionable to assert a mystical belief that numerical measurements are superior to other forms of understanding. Those who promote the use of citation statistics as a *replacement* for a fuller understanding of research implicitly hold such a belief. We not only need to use statistics *correctly*—we need to use them *wisely* as well.

We do not argue with the effort to evaluate research but rather with the demand that such evaluations rely predominantly on "simple and objective" citation-based metrics—a demand that often is interpreted as requiring easy-to-calculate numbers that rank publications or people or programs. Research usually has multiple goals, both short-term and long, and it is therefore reasonable that its value must be judged by multiple criteria. Mathematicians know that there are many things, both real and abstract, that cannot be simply ordered, in the sense that each two can be compared. Comparison often requires a more complicated analysis, which sometimes leaves one undecided about which of two things is "better." The correct answer to "Which is better?" is sometimes: "It depends!"

The plea to use multiple methods to assess the quality of research has been made before (e.g. Martin, 1996 or Carey, Cowling and Taylor, 2007). Publications can be judged in many ways, not only by citations. Measures of esteem such as invitations, membership on editorial boards, and awards often measure quality. In some disciplines and in some countries, grant funding can play a role. And peer review—the judgment of fellow scientists—is an important component of assessment. (We should not discard peer review merely because it is sometimes flawed by bias, any more than we should discard citation statistics because they are sometimes flawed by misuse.) This is a small sample of the multiple ways in which assessment can be done. There are many avenues to good assessment, and their relative importance varies among disciplines. In spite of this, "objective" citation-based statistics repeatedly become the preferred method for assessment. The lure of a simple process and simple numbers (preferably a single number) seems to overcome common sense and good judgment.

> Research usually has multiple goals and it is therefore reasonable that its value must be judged by multiple criteria.

This report is written by mathematical scientists to address the misuse of statistics in assessing scientific research. Of course, this misuse is sometimes directed towards the discipline of mathematics itself, and that is one of the reasons for writing this report. The special citation culture of mathematics, with low citation counts for journals, papers, and authors, makes it especially vulnerable to the abuse of citation statistics. We believe, however, that *all* scientists, as well as the general public, should be anxious to use sound scientific methods when assessing research.

Some in the scientific community would dispense with citation statistics altogether in a cynical reaction to past abuse, but doing so would mean discarding a valuable tool. Citation-based statistics *can* play a role in the assessment of research, provided they are used properly, interpreted with caution, and make up only part of the process. Citations provide information about journals, papers, and people.

We don't want to hide that information; we want to illuminate it.

That is the purpose of this report. The first three sections address the ways in which citation data can be used (and misused) to evaluate journals, papers, and people. The next section discusses the varied meanings of citations and the consequent limitations on citation-based statistics. The last section counsels about the wise use of statistics and urges that assessments temper the use of citation statistics with other judgments, even though it makes assessments less simple.

"Everything should be made as simple as possible, but not simpler," Albert Einstein once said.[1] This advice from one of the world's preeminent scientists is especially apt when assessing scientific research.

## RANKING JOURNALS: THE IMPACT FACTOR[2]

The impact factor was created in the 1960s as a way to measure the value of journals by calculating the average number of citations per article over a specific period of time (Garfield, 2005). The average is computed from data gathered by *Thomson Scientific* (previously called the Institute for Scientific Information), which publishes *Journal Citation Reports* (THOMSON: JOURNAL CITATION REPORTS). *Thomson Scientific* extracts references from more than 9000 journals, adding information about each article and its references to its database each year (THOMSON: SELECTION). Using that information, one can count how often a particular article is cited by subsequent articles that are published in the collection of indexed journals. (We note that *Thomson Scientific* indexes less than half the mathematics journals covered by *Mathematical Reviews* and *Zentralblatt*, the two major reviewing journals in mathematics.[3])

For a particular journal and year, the journal impact factor is computed by calculating the average number of citations to articles in the journal during the preceding two years from all articles published in that given year (in the particular collection of journals indexed by *Thomson Scientific*). If the impact factor of a journal is 1.5 in 2007, it means that on average articles published during 2005 and 2006 were cited 1.5 times by articles in the collection of all indexed journals published in 2007.

*Thomson Scientific* itself uses the impact factor as one factor in selecting which journals to index (THOMSON: SELECTION). On the other hand, Thomson promotes the use of the impact factor more generally to compare journals.

> "As a tool for management of library journal collections, the impact factor supplies the library administrator with information about journals already in the collection and journals under consideration for acquisition. These data must also be combined with cost and circulation data to make rational decisions about purchases of journals" (THOMSON: IMPACT FACTOR).

Many writers have pointed out that one should not judge the academic worth of a journal using citation data alone, and the present authors very much agree. In addition to this general observation, the impact factor has been criticized for other reasons as well. (See Seglen, 1997; Amin and Mabe, 2000;

---

[1] This quote was attributed to Einstein in the *Reader's Digest*. Oct. 1977. It appears to be derived from his actual quote: "It can scarcely be denied that the supreme goal of all theory is to make the irreducible basic elements as simple and as few as possible without having to surrender the adequate representation of a single datum of experience." From "On the Method of Theoretical Physics" The Herbert Spencer Lecture, delivered at Oxford (10 June 1933); also published in *Philosophy of Science* **1** 163–169.

[2] While we concentrate on the *Thomson Scientific* impact factor in this section, we note that Thomson promotes the use of two other statistics. Also, similar statistics based on average citation counts for journals can be derived from other databases, including Scopus, Spires, Google Scholar and (for mathematics) the Math Reviews citation database. The latter consists of citations from over 400 mathematics journals from the period 2000–present, identified as items that were listed in Math Reviews since 1940; it includes more than 3 million citations.

[3] *Thomson Scientific* indicates (March 2008) that it indexes journals in the following categories:

> MATHEMATICS (217),
> MATHEMATICS APPLIED (177),
> MATHEMATICS INTERDISCIPLINARY (76),
> PHYSICS, MATHEMATICAL (44),
> PROBABILITY AND STATISTICS (96).

The categories overlap, and the total number of journals is approximately 400. By contrast, *Mathematical Reviews* includes items from well more than 1200 journals each year, and considers more than 800 journals as "core" (in the sense that every item in the journal is included in Math Reviews). Zentralblatt covers a similar number of mathematics journals.

Monastersky, 2005; Ewing, 2006; Adler, 2007 and Hall, 2007.)

(i) The identification of the impact factor as an average is not quite correct. Because many journals publish non-substantive items such as letters or editorials, which are seldom cited, these items are not counted in the denominator of the impact factor. On the other hand, while infrequent, these items are sometimes cited, and these citations *are* counted in the numerator. The impact factor is therefore not quite the average citations per article. When journals publish a large number of such "non-substantial" items, this deviation can be significant. In many areas, including mathematics, this deviation is minimal.

(ii) The two-year period used in defining the impact factor was intended to make the statistic current (Garfield, 2005). For some fields, such as biomedical sciences, this is appropriate because most published articles receive most of their citations soon after publication. In other fields, such as mathematics, most citations occur beyond the two-year period. Examining a collection of more than 3 million recent citations in mathematics journals (the Math Reviews Citation database) one sees that roughly 90% of citations to a journal fall outside this 2-year window. Consequently, the impact factor is based on a mere 10% of the citation activity and misses the vast majority of citations.[4]

Does the two-year interval mean the impact factor is misleading? For mathematics journals the evidence is equivocal. *Thomson Scientific* computes 5-year impact factors, which it points out correlate well with the usual (2-year) impact factors (Garfield, 1998). Using the Math Reviews citation database, one can compute "impact factors" (i.e., average citations per article) for a collection of the 100 most cited mathematics journals using periods of 2, 5 and 10 years. The chart below shows that 5- and 10-year impact factors generally track the 2-year impact factor.

The one large outlier is a journal that did not publish papers during part of this time; the smaller outliers tend to be journals that publish a relatively small number of papers each year, and the chart merely reflects the normal variability in impact factors for such journals. It is apparent that changing the number of "target years" when calculating the impact factor changes the ranking of journals, but the changes are generally modest, except for small journals, where impact factors also vary when changing the "source year" (see below).

(iii) The impact factor varies considerably among disciplines (Amin and Mabe, 2000). Part of this difference stems from the observation (ii): If in some disciplines many citations occur outside the two-year window, impact factors for journals will be far lower. On the other hand, part of the difference is simply that the citation cultures differ from discipline to discipline, and scientists will cite papers at different rates and for different reasons. (We elaborate on this observation later because the meaning of citations is extremely important.) It follows that one cannot in any meaningful way compare two journals in different disciplines using impact factors.

(iv) The impact factor can vary considerably from year to year, and the variation tends to be larger for smaller journals (Amin and Mabe, 2000). For journals publishing fewer than 50 articles, for example, the average *change* in the impact factor from 2002 to 2003 was nearly 50%. This is wholly expected, of course, because the sample size for small journals is small. On the other hand, one often compares journals for a fixed year, without taking into account the higher variation for small journals.

(v) Journals that publish articles in languages other than English will likely receive fewer citations because a large portion of the scientific community cannot (or do not) read them. And the type of journal, rather than the quality alone, may influence the impact factor. Journals that publish review articles, for example, will often receive far more citations than journals that do not, and therefore have higher (sometimes, substantially higher) impact factors (Amin and Mabe, 2000).

(vi) The most important criticism of the impact factor is that its meaning is not well understood. When using the impact factor to compare two journals, there is no a priori model that defines what it means to be "better." The only model derives from the impact factor itself—a larger impact factor means a better journal. In the classical statistical paradigm, one defines a model, formulates a hypothesis (of no difference), and then finds a statistic, which depending on its values allows one to accept or reject the hypothesis. Deriving information (and possibly a model) from the data itself is a legitimate approach to statistical analysis, but in this case it is not clear what information has been derived. How does the impact factor measure quality? Is it the best statistic to measure quality? What precisely *does* it measure? (Our later discussion about the meaning of citations is relevant here.) Remarkably little is known about a model for journal quality or how it might relate to the impact factor.

---

[4]The *Mathematical Reviews* citation database includes (March 2008) more than 3 million references in approximately 400 journals published from 2000 to the present. The references are matched to items in the MR database and extend over many decades. Unlike the Science Citation Index, citations both to books and journals are included. It is a curious fact that roughly 50% of the citations are to items appearing in the previous decade; 25% cite articles appearing in the decade before that; 12.5% cite articles in the prior decade; and so on. This sort of behavior is special to each discipline, of course.

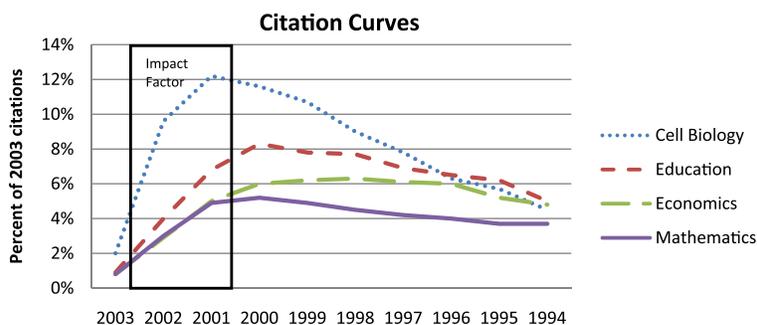

FIG. 1. *The age of citations from articles published in 2003 covering four different fields. Citations to articles published in 2001–2002 are those contributing to the impact factor; all other citations are irrelevant to the impact factor. Data from Thomson Scientific.*

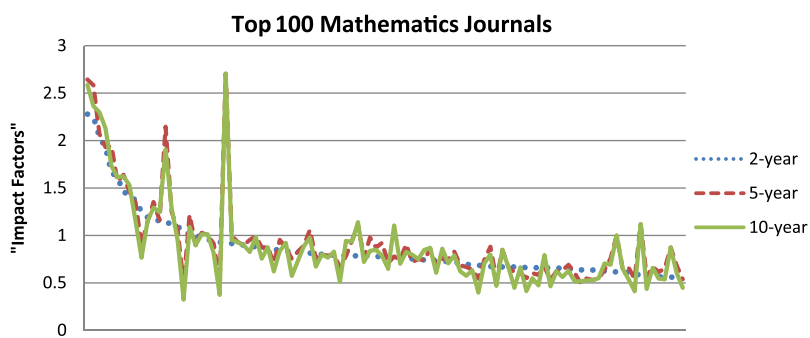

FIG. 2. *"Impact factors" for 2, 5 and 10 years for 100 mathematics journals. Data from Math Reviews citation database.*

The above six criticisms of the impact factor are all valid, but they mean only that the impact factor is crude, not useless. For example, the impact factor can be used as a starting point in ranking journals in groups by using impact factors initially to define the groups and then employing other criteria to refine the ranking and verify that the groups make sense. But using the impact factor to evaluate journals requires caution. The impact factor cannot be used to compare journals across disciplines, for example, and one must look closely at the type of journals when using the impact factor to rank them. One should also pay close attention to annual variations, especially for smaller journals, and understand that small differences may be purely random phenomena. And it is important to recognize that the impact factor may not accurately reflect the full range of citation activity in some disciplines, both because not all journals are indexed and because the time period is too short. Other statistics based on longer periods of time and more journals may be better indicators of quality. Finally, citations are only one way to judge journals, and should be supplemented with other information (the central message of this report).

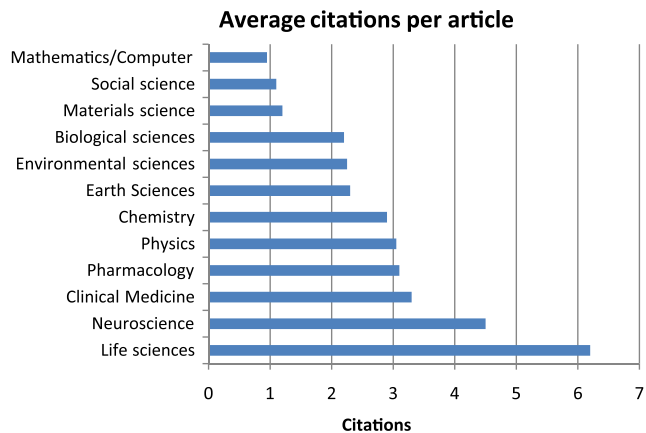

FIG. 3. *Average citations per article for different disciplines, showing that citation practices differ markedly. Data from Thomson Scientific (Amin and Mabe, 2000).*

These are all cautions similar to those one would make for any ranking based on statistics. Mindlessly ranking journals according to impact factors for a particular year is a misuse of statistics. To its credit, *Thomson Scientific* agrees with this statement and (gently) cautions those who use the impact factor about these things.

> "Thomson Scientific does not depend on the impact factor alone in assessing the usefulness of a journal, and neither should anyone else. The impact factor should not be used without careful attention to the many phenomena that influence citation rates, as for example the average number of references cited in the average article. The impact factor should be used with informed peer review" (THOMSON: IMPACT FACTOR).

Unfortunately, this advice is too often ignored.

## RANKING PAPERS

The impact factor and similar citation-based statistics can be misused when ranking journals, but there is a more fundamental and more insidious misuse: Using the impact factor to compare individual papers, people, programs, or even disciplines. This is a growing problem that extends across many nations and many disciplines, made worse by recent national research assessments.

In a sense, this is not a new phenomenon. Scientists are often called upon to make judgments about publication records, and one hears comments such as, "She publishes in good journals" or "Most of his papers are in low level journals." These can be sensible assessments: The quality of journals in which a scientist generally (or consistently) publishes is one of many factors one can use to assess the scientist's overall research. The impact factor, however, has increased the tendency to ascribe the properties of an individual journal to *each* article within that journal (and to *each* author).

*Thomson Scientific* implicitly promotes this practice:

> "Perhaps the most important and recent use of impact is in the process of academic evaluation. The impact factor can be used to provide a gross approximation of the prestige of journals in which individuals have been published" (THOMSON: IMPACT FACTOR).

Here are some examples of the ways in which people have interpreted this advice, reported from mathematicians around the world:

EXAMPLE 1. My university has recently introduced a new classification of journals using the Science Citation Index Core journals. The journals are divided into three groups based only on the impact factor. There are 30 journals in the top list, containing no mathematics journal. The second list contains 667, which includes 21 mathematics journals. Publication in the first list causes university support of research to triple; publication in the second list, to double. Publication in the core list awards 15 points; publication in any *Thomson Scientific* covered journal awards 10. Promotion requires a fixed minimum number of points.

EXAMPLE 2. In my country, university faculty with permanent positions are evaluated every six years. Sequential successful evaluations are the key to all academic success. In addition to a curriculum vitae, the largest factor in evaluation concerns ranking five published papers. In recent years, these are given 3 points if they appear in journals in the top third of the *Thomson Scientific* list, 2 points if in the second third, and 1 point in the bottom third. (The three lists are created using the impact factor.)

EXAMPLE 3. In our department, each faculty member is evaluated by a formula involving the number of single-author-equivalent papers, multiplied by the impact factor of the journals in which they appear. Promotions and hiring are based partly on this formula.

In these examples, as well as many others reported to us, the impact factor is being used either explicitly or implicitly to compare individual papers along with their authors: If the impact factor of journal A is greater than that of journal B, then surely a paper in A must be superior to a paper in B, and author

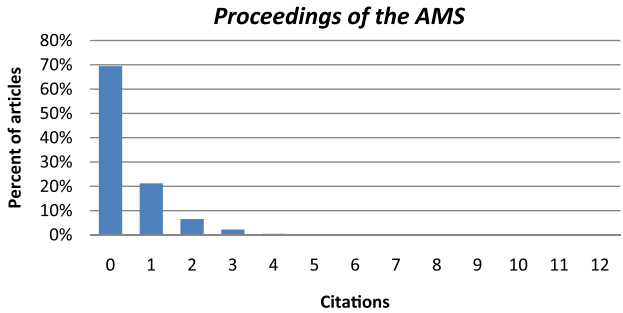

Fig. 4. *Citation distribution for papers in Proceedings of the American Mathematical Society, 2000–2004.*

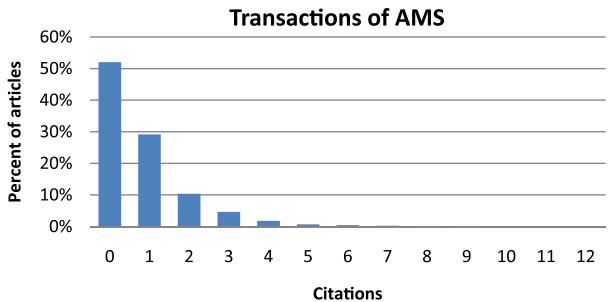

Fig. 5. *Citation distribution for papers in the Transactions of the American Mathematical Society, 2000–2004.*

A superior to author B. In some cases, this reasoning is extended to rank departments or even entire disciplines.

It has long been known that the distribution of citation counts for individual papers in a journal is highly skewed, approximating a so-called power law (Seglen, 1997; Garfield, 1987). This has consequences that can be made precise with an example.

The distribution for papers in the *Proceedings of the American Mathematical Society* over the period 2000–2004 can be seen below. The *Proceedings* publishes short papers, normally shorter than ten pages in length. During this period, it published 2381 papers (about 15,000 pages). Using 2005 journals in the Math Reviews citation database, the average citation count per article (that is, the impact factor) is 0.434.

The *Transactions of the AMS* publishes longer articles that are usually more substantial, both in scope and content. Over the same period of time, the *Transactions* published 1165 papers (more than 25,000 pages), with citation counts ranging from 0 to 12. The average number of citations per article was 0.846—about twice that of the *Proceedings*.

Now consider two mathematicians, one publishing a paper in the Proceedings and the other a paper in the Transactions. Using some of the institutional practices cited above, the second would be judged superior to the first, publishing a paper in a journal with higher impact factor—in fact, twice as high! Is this a valid assessment? Are papers in the *Transaction of the AMS* twice as good as those in the *Proceedings*?

When we assert that an individual *Transactions* paper is better (in the sense of citations) than an individual *Proceedings* paper, we need to ask not a question about averages, but rather a question about probabilities: What is the probability that we are wrong? What is the probability that a randomly selected *Proceedings* paper has at least as many citations as a randomly selected *Transactions* paper?

This is an elementary calculation, and the answer is 62%. This means that we are wrong 62% of the time, and a randomly selected *Proceedings* paper will be just as good as (or better than) a randomly selected *Transactions* paper—in spite of the fact that the *Proceedings* impact factor is only half that of the *Transactions*! We are more often wrong than right. *Most* people find this surprising, but it is a consequence of the highly skewed distribution and the narrow window of time used to compute the impact factor (which is the reason for the high percentage of uncited papers).[5] It shows the value of precise statistical thinking rather than intuitive observation.

This is typical behavior for journals, and there is nothing special about the choices of these two journals. (For example, the *Journal of the AMS* over the same period has an impact factor 2.63—six times

---

[5]The skewed distribution combined with the narrow window (using only one year's journals as the source of citations and five years as the target) means that a large number of articles have either none or very few citations. This makes it intuitively obvious that randomly chosen articles are often equivalent.

The fact that many articles have no citations (or only a few) is also a consequence of the long citation time for mathematics—articles often take many years to accumulate citations. If we choose longer periods of time for both source journals and target years, then the citation counts increase substantially and it becomes easier to distinguish journals by citation behavior. This is the approach used in (Stringer et al., 2008) to analyze citations. They show that for sufficiently long periods of time, the distribution of citation counts for individual articles appears to be log-normal. This provides a mechanism for comparing two journals by comparing the distributions, and is certainly more sophisticated than using impact factors. Again, however, it considers only citations and nothing else.

that of the *Proceedings*. Yet a randomly selected *Proceedings* article is at least as good as a *Journal* article, in the sense of citations, 32% of the time.)

Thus, while it is incorrect to say that the impact factor gives no information about individual papers in a journal, the information is surprisingly vague and can be dramatically misleading.

It follows that the kinds of calculations performed in the three examples above—using the impact factor as a proxy for actual citation counts for individual papers—have little rational basis. Making assertions that are incorrect more than half the time (or a third of the time) is surely not a good way to carry out an assessment.

Once one realizes that it makes no sense to substitute the impact factor for individual article citation counts, it follows that it makes no sense to use the impact factor to evaluate the authors of those articles, the programs in which they work, and (most certainly) the disciplines they represent. The impact factor and averages in general are too crude to make sensible comparisons of this sort without more information.

Of course, ranking people is not the same as ranking their papers. But if you want to rank a person's papers using only citations to measure the quality of a particular paper, you must begin by counting that paper's citations. The impact factor of the journal in which the paper appears is not a reliable substitute.

> While it is incorrect to say that the impact factor gives no information about individual papers in a journal, the information is surprisingly vague and can be dramatically misleading.

## RANKING SCIENTISTS

While the impact factor has been the best known citation-based statistic, there are other more recent statistics that are now actively promoted. Here is a small sample of three of these statistics meant to rank individuals.

*h-index*: A scientist's h-index is the largest $n$ for which he/she has published $n$ articles, each with at least $n$ citations.

This is the most popular of the statistics mentioned here. It was proposed by J. E. Hirsch (Hirsch, 2006) in order to measure "the scientific output of a researcher" by focusing on the high-end "tail" of a person's citation distribution. The goal was to substitute a single number for publications counts and citation distributions.

*m-index*: A scientist's m-index is the h-index divided by the number of years since his/her first paper.

This was also proposed by Hirsch in the paper above. The intention is to compensate junior scientists because they have not had time to publish papers or gain many citations.

*g-index*: A scientist's g-index is the largest $n$ for which the $n$ most cited papers have a total of at least $n^2$ citations.

This was proposed by Leo Egghe in 2006 (Egghe, 2006). The h-index does not take into account the fact that some papers in the top $n$ may have extraordinarily high citation counts. The g-index is meant to compensate for this.

There are more indices—many more of them—including variants of those above that take into account the age of papers or the number of authors (Batista et al., 2005; Batista, Campiteli and Kinouchi, 2006; Sidiropouls, Katsaros and Manolopoulos, 2006).

In his paper defining the h-index, Hirsch wrote that he proposed the h-index as "an easily computable index, which gives an estimate of the importance, significance, and broad impact of a scientist's cumulative research contributions" (Hirsch, 2006). He went on to add that "this index may provide a useful yardstick to compare different individuals competing for the same resource when an important evaluation criterion is scientific achievement."

Neither of these assertions is supported by convincing evidence. To support his claim that the h-index measures the importance and significance of a scientist's cumulative research, Hirsch analyzes the h-index for a collection of Nobel Prize winners (and, separately, members of the National Academy). He demonstrates that people in these groups generally have high h-indices. One can conclude that it is likely a scientist has a high h-index given the scientist is a Nobel Laureate. But without further information, we know very little about the likelihood someone will become a Nobel Laureate or a member of the National Academy, given that they have a high h-index. That is the kind of information one wants in order to establish the validity of the h-index.

In his article, Hirsch also claims that one can use the h-index to compare two scientists:

"I argue that two individuals with similar h are comparable in terms of their overall scientific impact, even if their total number of papers or their total number of citations is very different. Conversely, that between two individuals (of the same scientific age) with similar number of total papers or of total citation count and very different h-value, the one with the higher h is likely to be the more accomplished scientist" (Hirsch, 2006).

These assertions appear to be refuted by common sense. (Think of two scientists, each with 10 papers with 10 citations, but one with an additional 90 papers with 9 citations each; or suppose one has exactly 10 papers of 10 citations and the other exactly 10 papers of 100 each. Would anyone think them equivalent?)[6]

Hirsch extols the virtues of the h-index by claiming that "h is preferable to other single-number criteria commonly used to evaluate scientific output of a researcher..." (Hirsch, 2006), but he neither defines "preferable" nor explains why one *wants* to find "single-number criteria."

While there has been some criticism of this approach, there has been little serious analysis. Much of the analysis consists of showing "convergent validity," that is, the h-index correlates well with other publication/citation metrics, such as the number of published papers or the total number of citations. This correlation is unremarkable, since all these variables are functions of the same basic phenomenon—publications. In one notable paper about the h-index (Lehmann, Jackson and Lautrup, 2006) the authors carry out a more careful analysis and demonstrate that the h-index (actually, the m-index) is not as "good" as merely considering the mean number of citations per paper. Even here, however, the authors do not adequately define what the term "good" means. When the classical statistical paradigm is applied (Lehmann, Jackson and Lautrup, 2006), the h-index proves to be less reliable than other measures.

A number of variants of the h-index have been devised to compare the quality of researchers not only within a discipline but across disciplines as well (Batista, Campiteli and Kinouchi, 2006; Molinari and Molinari, 2008). Others claim that the h-index can be used to compare institutes and departments (Kinney, 2007). These are often breathtakingly naïve attempts to capture a complex citation record with a single number. Indeed, the primary advantage of these new indices over simple histograms of citation counts is that the indices discard almost all the detail of citation records, and this makes it possible to rank any two scientists. Even simple examples, however, show that the discarded information is needed to understand a research record. Surely understanding ought to be the goal when assessing research, not merely ensuring that any two people are comparable.

> Understanding ought to be the goal when assessing research, not merely ensuring that any two people are comparable.

In some cases, national assessment bodies are gathering the h-index or one of its variants as part of their data. This is a misuse of the data. Unfortunately, having a single number to rank each scientist is a seductive notion—one that may spread more broadly to a public that often misunderstands the proper use of statistical reasoning in far simpler settings.

---

[6]To illustrate how much information one loses when using only the h-index, here is a real-life example of a distinguished mid-career mathematician who has published 84 research papers. The citation distribution looks like the following:

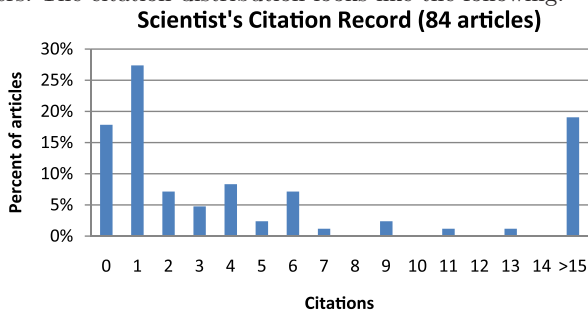

Notice that slightly under 20% of the publications have 15 or more citations. The distribution of actual citation counts for these 15 papers is:

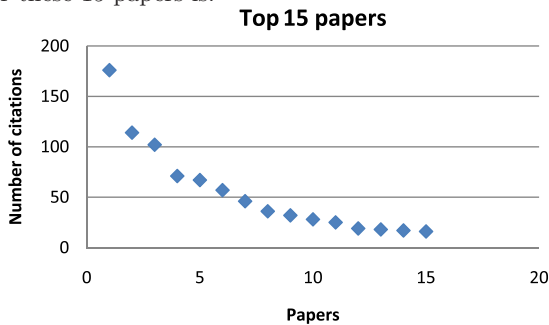

In Hirsch's analysis, however, all this information is thrown away. One only remembers that the h-index is 15, meaning that the top 15 papers have 15 or more citations.

## THE MEANING OF CITATIONS

Those who promote citation statistics as the predominant measure of research quality do not answer the essential question: What do citations mean? They gather large amounts of data about citation counts, process the data in order to derive statistics, and then assert the resulting assessment process is "objective." Yet it is the *interpretation* of the statistics that leads to assessment, and the interpretation relies on the *meaning* of citations, which is quite subjective.

In the literature promoting this approach, it is surprisingly difficult to find clear statements about the meaning of citations.

> "The concept behind citation indexing is fundamentally simple. By recognizing that the value of information is determined by those who use it, what better way to measure the quality of the work than by measuring the impact it makes on the community at large. The widest possible population within the scholarly community (i.e. anyone who uses or cites the source material) determines the influence or impact of the idea and its originator on our body of knowledge" (THOMSON: HISTORY).

> "Although quantifying the quality of individual scientists is difficult, the general view is that it is better to publish more than less and that the citation count of a paper (relative to citation habits in the field) is a useful measure of it quality" (Lehman, Jackson and Lautrup 2006, page 1003).

> "Citation frequency reflects a journal's value and the use made of it..."

> "When a physician or a biomedical researcher cites a journal article, it indicates that the cited journal has influenced him or her in some manner" (Garfield, 1987, page 7).

> "Citations are an acknowledgement of intellectual debt" (THOMSON: FIFTY YEARS).

The relevant terms are "quality," "value," "influence" and "intellectual debt." The term "impact" has become the generic word used to assign meaning to citations—a term that first arose in a short paper written in 1955 by Eugene Garfield to promote the idea of creating a citation index. He wrote:

> "Thus, in the case of a highly significant article, the citation index has a quantitative value, for it may help the historian to measure the influence of the article—that is, its 'impact factor'" (Garfield, 1955, page 3).

It is fairly clear that here, as elsewhere, the term "impact factor" is intended to suggest that the citing paper has been "built upon" the work of the cited—that citations are the mechanism by which research propagates itself forward.

There is a rich literature about the actual meaning of citations that suggests citations are more complicated than these vague statements lead us to believe. For example, in their 1983 paper on assessing research, Martin and Irvine write:

> "Underlying all these problems with the use of citations as a measure of quality is our ignorance of the reasons *why* authors cite particular pieces of work and not others. The problems described above... Simple citation analysis presupposes a highly rational model of reference-giving, in which citations are held to reflect primarily scientific appreciation of previous work of high quality or importance, and potential citers all have the same chance to cite particular papers..." (Martin and Irvine, 1983, page 69).

In her 1988 paper on the meaning of citations (Cozzens, 1989), Cozzens asserts that citations are the result of two systems underlying the conduct of scientific publication, one a "reward" system and the other "rhetorical." The first kind have the meaning most often associated with a citation—an acknowledgment that the citing paper has "intellectual debt" to the cited. The second, however, have a meaning quite different—a reference to a previous paper that explains some result, perhaps not a result of the cited author at all. Such rhetorical citations are merely a way to carry on a scientific conversation, not establish intellectual indebtedness. Of course, in some cases, a citation can have both meanings.

> The meaning of citations is not simple and citation-based statistics are not nearly as "objective" as proponents assert.

Cozzens makes the observation that *most* citations are rhetorical. This is confirmed by the experience of most practicing mathematicians. (In the Math Reviews citations database, for example, nearly 30% of the more than 3 million citations are to books and *not* to research articles in journals.) Why is this important? Because unlike "reward" citations, which tend to refer to seminal papers, the choice of which paper to cite rhetorically depends on many factors—the prestige of the cited author (the "halo" effect), the relationship of the citing and cited authors, the availability of the journal (Are open access journals more likely to be cited?), the convenience of referencing several results from a single paper, and so forth. Few of these factors are directly related to the "quality" of the cited paper.

Even when citations are "reward" citations, they can reflect a variety of motives, including "currency, negative credit, operational Information, persuasiveness, positive credit, reader alert, and social consensus" (Brooks, 1986). In most cases, citations were motivated by more than one of these. Some notable results can suffer the "obliteration" effect, immediately being incorporated into the work of others, which then serves as the basis for further citations. Other citations are not rewards for outstanding research, but rather warnings about flawed results or thinking. The present report provides many examples of such "warning" citations.

The sociology of citations is a complex subject—one that is beyond the scope of this report. Even this cursory discussion, however, shows that the meaning of citations is not simple and that citation-based statistics are not nearly as "objective" as proponents assert.

Some might argue that the meaning of citations is immaterial because citation-based statistics are highly correlated with some other measure of research quality (such as peer review). For example, the *Evidence* report mentioned earlier argues that citation-statistics can (and should) replace other forms of evaluation because of this correlation:

> "Evidence has argued that bibliometric techniques can create indicators of research quality that are congruent with researcher perception" (Evidence Report, 2007, page 9).

The conclusion seems to be that citation-based statistics, regardless of their precise meaning, should replace other methods of assessment, because they often agree with them. Aside from the circularity of this argument, the fallacy of such reasoning is easy to see.

## USING STATISTICS WISELY

The zealous over-reliance on objective metrics (statistics) to assess research is neither a new nor an isolated phenomenon. It is eloquently described in the 2001 popular book, *Damned lies and statistics*, written by the sociologist Joel Best:

> "There are cultures in which people believe that some objects have magical powers; anthropologists call these objects fetishes. In our society, statistics are a sort of fetish. We tend to regard statistics as though they are magical, as though they are more than mere numbers. We treat them as powerful representations of the truth; we act as though they distill the complexity and confusion of reality into simple facts. We use statistics to convert complicated social problems into more easily understood estimates, percentages, and rates. Statistics direct our concern; they show us what we ought to worry about and how much we ought to worry. In a sense, the social problem becomes the statistic and, because we treat statistics as true and incontrovertible, they achieve a kind of fetish-like, magical control over how we view social problems. We think of statistics as facts that we discover, not numbers we create" (Best, 2001, page 160).

This mystical belief in the magic of citation statistics can be found throughout the documentation for research assessment exercises, both national and institutional. It can also be found in the work of those promoting the h-index and its variants.

This attitude is also evident in recent attempts to improve on the impact factor using more sophisticated mathematical algorithms, including page rank algorithms, to analyze citations (Bergstrom, 2007; Stringer et al., 2008). Their proponents make claims about their efficacy that are unjustified by the analysis and difficult to assess. Because they are based on more complicated calculations, the (often hidden) assumptions behind them are not easy for most people to discern.[7] We are meant to treat the num-

---

[7]The algorithm in (Bergstrom, 2007) uses a page rank algorithm to give each citation a weight, and then computes an

bers and rankings with awe—as truths rather than creations.

Research is not the first publicly funded activity to come under scrutiny, and over the past decades people have tried to carry out quantitative performance assessments of everything from educational systems (schools) to healthcare (hospitals and even individual surgeons). In some cases, statisticians have stepped in to advise those doing the measuring about sensible metrics and the proper use of statistics. If one consults with doctors when practicing medicine, surely one ought to consult with (and heed the advice of) statisticians when practicing statistics. Two excellent examples can be found in (Bird et al., 2005) and (Goldstein and Spiegelhalter, 1996). While they each deal with performance assessment of things other than research—public sector performance monitoring in the first and healthcare/education in the second—each provides insight about the sensible use of statistics in assessing research.

> If one consults with doctors when practicing medicine, surely one ought to consult with statisticians when practicing statistics.

The paper by Goldstein and Spiegelhalter in particular deals with the use of League Tables (rankings) based on simple numbers (e.g., student achievements or medical outcomes), and it is particularly relevant to assessing research by ranking journals, papers, or authors using citation statistics. In their paper, the authors outline a three-part framework for any performance assessment:

### Data

> "No amount of fancy statistical footwork will overcome basic inadequacies in either the *appropriateness* or the *integrity* of the data collected" (Goldstein and Spiegelhalter, 1996, page 389).

This is an important observation for citation-based performance assessment. The impact factor, for example, is based on a subset of data, which includes only those journals selected by *Thomson Scientific*. (We note that the impact factor itself is the major part of the selection criterion.) Some have questioned the integrity of this data (Rossner, Van Epps and Hill, 2007). Others point out that other data sets might be more complete (Meho and Yang, 2007). Several groups have pushed the idea of using Google Scholar to implement citation-based statistics, such as the h-index, but the data contained in Google Scholar is often inaccurate (since things like author names are automatically extracted from web postings). Citation statistics for individual scientists are sometimes difficult to obtain because authors are not uniquely identified, and in some settings and certain countries, this can be an enormous impediment to assembling accurate citation data. The particular collection of data one uses for citation analysis is frequently overlooked. One is likely to draw faulty conclusions from statistics based on faulty data.

### Statistical Analysis and Presentation

> "We shall pay particular attention to the specification of an appropriate statistical *model*, the crucial importance of *uncertainty* in the presentation of all results, techniques for *adjustment* of outcomes for confounding factors and finally the extent to which any reliance may be placed on explicit *rankings*" (Goldstein and Spiegelhalter, 1996, page 390).

As we have written previously, in most cases in which citation statistics are used to rank papers, people, and programs, no specific model is specified in advance. Instead, the data itself suggests a model, which is often vague. A circular process seems to rank objects higher because they are ranked higher (in the database). There is frequently scant attention to uncertainty in *any* of these rankings, and little analysis of how that uncertainty (e.g., annual variations in the impact factor) would affect the

---

"impact factor" by using the weighted averages for citations. Page rank algorithms have merit because they take into account the "value" of citations. On the other hand, their complexity can be dangerous because the final results are harder to understand. In this case, all "self-citations" are discarded—that is, all citations from articles in a given journal J to articles published in J during the preceding five years are discarded. These are not "self-citations" in any normal sense of the word, and a glance at some data from the Math Reviews Citations database suggests that this discards roughly one-third of all citations.

The algorithm in (Stringer et al., 2008) is interesting, in part because it attempts to address the differing time-scales for citations as well as the issue of comparing randomly selected papers in one journal with those from another. Again, the complexity of the algorithms makes it hard for most people to evaluate their results. One notable hypothesis is slipped into the paper on page 2: "Our first assumption is that the papers published in journal J have a normal distribution of 'quality'..." This seems to contradict common experience.

rankings. Finally, confounding factors (e.g., the particular discipline, the type of articles a journal publishes, whether a particular scientist is an experimentalist or theoretician) are frequently ignored in such rankings, especially when carried out in national performance assessments.

### Interpretation and Impact

> "The comparisons discussed in this paper are of great public interest, and this is clearly an area where careful attention to limitations is both vital and likely to be ignored. Whether adjusted outcomes are in any way valid measures of institutional 'quality' is one issue, while analysts should also be aware of the potential effect of the results in terms of future behavioural changes by institutions and individuals seeking to improve their subsequent 'ranking'" (Goldstein and Spiegelhalter, 1996, page 390).

The assessment of research is *also* of great public interest. For an individual scientist, an assessment can have profound and long-term effects on one's career; for a department, it can change prospects for success far into the future; for disciplines, a collection of assessments can make the difference between thriving and languishing. For a task so important, surely one should understand both the validity and the limitations of the tools being used to carry it out. To what extent do citations measure the quality of research? Citation counts seem to be correlated with quality, and there is an intuitive understanding that high-quality articles are highly-cited. But as explained above, some articles, especially in some disciplines, are highly-cited for reasons other than high quality, and it does not follow that highly-cited articles are necessarily high quality. The precise interpretation of rankings based on citation statistics needs to be better understood. In addition, if citation statistics play a central role in research assessment, it is clear that authors, editors and even publishers will find ways to manipulate the system to their advantage (Macdonald and Kam, 2007). The long-term implications of this are unclear and unstudied.

The article by Goldstein and Spiegelhalter is valuable to read today because it makes clear that the over-reliance on simple-minded statistics in research assessment is not an isolated problem. Governments, institutions, and individuals have struggled with similar problems in the past in other contexts, and they have found ways to better understand the statistical tools and to augment them with other means of assessment. Goldstein and Spiegelhalter end their paper with a positive statement of hope:

> "Finally, although we have been generally critical of many current attempts to provide judgments about institutions, we do not wish to give the impression that we believe that all such comparisons are necessarily flawed. It seems to us that the comparison of institutions and the attempt to understand why institutions differ is an extremely important activity and is best carried out in a spirit of collaboration rather than confrontation. It is perhaps the only sure method for obtaining objectively based information which can lead to understanding and ultimately result in improvements. *The real problem with the simplistic procedures which we have set out to criticize is that they distract both attention and resources from this worthier aim*" (Goldstein and Spiegelhalter, 1996, page 406).

It would be hard to find a better statement to express the goals that should be shared by everyone involved in the assessment of research.